\documentclass[preprint,aps,12pt,preprintnumbers,nofootinbib,superscriptaddress]
{revtex4}

\usepackage{graphicx}
\usepackage{amssymb,amsmath}
\def\lag{{\mathcal{L}}}

\DeclareMathOperator{\tr}{tr}

\newcommand\beq{\begin{equation}}
\newcommand\eeq{\end{equation}}

\begin{document}

\newcount\hour \newcount\minute
\hour=\time \divide \hour by 60
\minute=\time
\count99=\hour \multiply \count99 by -60 \advance \minute by \count99
\newcommand{\mydate}{\ \today \ - \number\hour :\number\minute}

\preprint{CALT 68-2662}
\preprint{UCSD/PTH 07-10}

\title{Massive Vector Scattering in Lee-Wick Gauge Theory}
\author{Benjam\'in Grinstein}
\email[]{bgrinstein@ucsd.edu}
\affiliation{Department of Physics, University of California at San Diego, La Jolla, CA 92093}

\author{Donal O'Connell}
\email[]{donal@ias.edu}
\affiliation{Institute for Advanced Study, School of Natural Sciences, Einstein Drive, Princeton, NJ 08540}
\affiliation{The Niels Bohr Institute, Blegdamsvej 17, DK-2100 Copenhagen, Denmark}

\author{Mark B. Wise}
\email[]{wise@theory.caltech.edu}
\affiliation{California Institute of Technology, Pasadena, CA 91125}

\date{\today}

\begin{abstract}

We demonstrate that amplitudes describing scattering of longitudinally
polarized massive vector bosons present in non-Abelian Lee-Wick gauge
theory do not grow with energy and, hence, satisfy the constraints imposed 
by perturbative unitarity. This result contrasts with the widely-known violation 
of perturbative unitarity in the standard model with a very heavy Higgs. 
%We prove
Our conclusions are valid to all orders of perturbation theory and depend
%that unitarity obtains perturbatively to all orders in Lee-Wick gauge
%theories. Our results depend 
on the existence of a formulation of the
theory in which all operators are of dimension four or less. This can be
thought of as a restriction on the kinds of higher dimension operator
which can be included in the higher derivative formulation of the theory.

%Longitudinally polarized massive vector boson scattering amplitudes
%may violate perturbative unitarity due to the growth of the polarization
%vector with energy. Since such massive vectors are present in the Lee-Wick
%standard model, or in any non-Abelian Lee-Wick gauge theory, it may appear
%that these theories must become strongly coupled near the scale $M$
%associated with the Lee-Wick particles. However, we demonstrate to all
%orders of perturbation theory that the scattering amplitudes in Lee-Wick
%gauge theories do not grow with energy so that perturbative unitarity
%may be maintained. These results depend on the existence of the Lee-Wick
%form of the theory, in which all operators are of dimension 4 or less.
%This result is a consequence of the exact gauge symmetry of
%Lee-Wick gauge theories in addition to the existence of the Lee-Wick
%form of the Lagrangian in which all operators are of dimension 4 or less.

\end{abstract}

\maketitle

\section{Introduction}

Non-Abelian Lee-Wick gauge theories were introduced in reference~\cite{us},
building on pioneering work of Lee and Wick~\cite{leewick}. Lee and
Wick introduced a finite theory of quantum electrodynamics, which 
includes extra degrees of freedom that cancel radiative divergences
present in QED. These new degrees of freedom are associated with a 
non-positive definite norm on the Hilbert space. Lee and Wick argued
that the theory could nevertheless be unitary provided that the
new Lee-Wick particles obtain decay widths. Another peculiar feature of
the Lee-Wick theory is that it is classically unstable, so that a future
boundary condition must be imposed to prevent exponential growth of some
modes. This
leads to causality violation in the theory~\cite{Coleman}; however, this acausality is
suppressed below the scales associated with the Lee-Wick particles.

There has been considerable discussion of this proposal
in the literature, including debate about unitarity of the
theory~\cite{CLOP,Nakanishi:1971jj, Lee:1971ix, Nakanishi:1971ky},
possible applications to gravity~\cite{Antoniadis:1986tu},
as well as non-perturbative studies of Lee-Wick
theories~\cite{Kuti,Boulware:1983vw}. In~\cite{us} a non-Abelian extension
of the original proposal of Lee and Wick was described. Non-Abelian
Lee-Wick gauge theories are not finite but the only divergences present
are logarithmic. It was also shown in~\cite{us} how to build theories
with Lee-Wick partners of chiral fermions.  Consequently, it is possible
to write down an extension of the standard model to include Lee-Wick
fields. This Lee-Wick standard model has a stable Higgs mass and is
consistent with current observations if the Lee-Wick particles present
in the theory have masses of order the TeV scale.  Recently, there
has been further discussion of the Lee-Wick standard model, including
aspects of the LHC phenomenology of the model~\cite{Rizzo:2007ae,
Krauss:2007bz}, automatic suppression of flavor changing neutral currents
in the Lee-Wick standard model~\cite{Tim}, gravitational Lee-Wick
particles~\cite{Wu:2007yd}, and the possibility of coupling heavy physics
to the model~\cite{Espinosa:2007ny}.

In this work, we focus on an important conceptual issue in Lee-Wick
gauge theory.  These theories can be thought of as ordinary gauge theories
with additional higher dimension operators present in the Lagrangian. A
particular choice of higher dimension operator was made in~\cite{us}
but this was not the most general choice. The higher dimension operator
leads to the presence of interacting massive (Lee-Wick) vector bosons
in the theory. It was shown in~\cite{us} that an equivalent form of
the theory exists in which all operators are of dimension four or
less. In this formulation, new fields are present which describe the
Lee-Wick particles. One might think that scattering of massive Lee-Wick
vector bosons will violate perturbative unitarity at some scale since
their longitudinal polarizations grow with energy.  
%We will demonstrate
%that this does not occur in Lee-Wick gauge theory. 
We will demonstrate that this does not occur precisely for the very
special choice of higher derivative operators that can be written in
the form of a Lee-Wick gauge theory.
The existence of the
Lee-Wick form of the Lagrangian will be a crucial ingredient in our proof.
Thus, a gauge theory with arbitrary higher dimension operators added
is not unitary in perturbation theory. Only specific higher dimension
operators are consistent with unitary scattering. These operators are
such that a Lee-Wick form of the Lagrangian exists where only operators
with dimension less than or equal to four are present.

\section{Non-Abelian Lee-Wick Gauge Theory
\label{sec:review}}

In this section we review the construction of non-Abelian Lee-Wick gauge theories. The
Lagrangian is 
\begin{equation}
\lag_\mathrm{hd} =  - \frac{1}{2} \tr \hat{F}_{\mu \nu} \hat{F}^{\mu\nu} + \frac{1}{M^2} 
\tr \left( \hat D^{\mu} \hat F_{\mu\nu} \right) \left( \hat D^\lambda \hat F_\lambda{}^\nu \right) ,
\label{eq:gbHDlag}
\end{equation}
where $\hat F_{\mu \nu} = \partial_\mu \hat A_\nu - \partial_\nu \hat
A_\mu - i g [ \hat A_\mu, \hat A_\nu]$, and $\hat A_\mu
= \hat A^A_\mu T^A$ with $T^A$ the generators of the gauge group $G$ in the
fundamental representation. We will refer to this as the higher derivative formulation of
the theory. We can derive an equivalent formulation as follows. First, we introduce an 
auxiliary vector field $\tilde A_\mu$ so that we can write the Lagrangian of the theory as
\begin{equation}
\lag = - \frac{1}{2} \tr \hat{F}_{\mu \nu} \hat{F}^{\mu\nu} - M^2 \tr \tilde A_\mu \tilde A^\mu + 2 \tr \hat F_{\mu\nu} \hat D^\mu \tilde A^\nu ,
\label{eq:gbLWlagNotDiag}
\end{equation}
where $\hat D_\mu \tilde A_\nu =
\partial_\mu \tilde A_\nu - i g [\hat A_\mu, \tilde A_\nu]$.
To diagonalize the kinetic terms, we introduce shifted fields defined by
\begin{equation}
\hat A_\mu = A_\mu + \tilde A_\mu.
\end{equation}
The Lagrangian becomes
\begin{multline}
\lag_\mathrm{LW} = - \frac{1}{2} \tr F_{\mu \nu} F^{\mu \nu} + \frac{1}{2} \tr \left( D_\mu \tilde A_\nu - D_\nu \tilde A_\mu \right)\left(D^\mu \tilde A^\nu - D^\nu \tilde A^\mu \right) 
-i g \tr \left( \left[ \tilde A_\mu, \tilde A_\nu \right] F^{\mu \nu} \right) 
\\
-\frac{3}{2} g^2 \tr \left( \left[ \tilde A_\mu, \tilde A_\nu \right]  \left[ \tilde A^\mu, \tilde A^\nu \right] \right) - 4 i g \tr \left( \left[ \tilde A_\mu, \tilde A_\nu \right] D^\mu \tilde A^\nu \right) - M^2 \tr \left(\tilde A_\mu \tilde A^\mu \right) .
\label{eq:gbLWlag}
\end{multline}
Note that in this (Lee-Wick) formulation only dimension four operators
appear in the Lagrangian. It is also evident in this form that the
theory is unstable because of the wrong sign kinetic terms for the field
$\tilde A_\mu$. We impose a future boundary condition that there is no
exponential growth of any mode to deal with this instability. 
%At the
%quantum level, further adjustments to the theory are necessary to ensure
%that the only cuts present in loop graphs are associated with physical
%states. 
%It is not our purpose here to discuss these cuts, see~\cite{us}
%for more information. 

The Lagrangian given in Eq.~\eqref{eq:gbLWlag} contains an interacting
massive Lee-Wick vector boson $\tilde A_\mu$.  These massive vectors
obtain widths since they can decay to ordinary gauge bosons. This
width is necessary to remove unphysical cuts in Feynman diagrams
associated with single Lee-Wick particles which would otherwise violate
unitarity\footnote{Further adjustments to the theory are necessary to
ensure that the only cuts present in loop graphs are associated with
physical states; for a pedagogical discussion of this topic, see~\cite{Coleman}.}. 
In this paper, we focus on constraints unitarity places on
the growth of amplitudes with energy and so we neglect the widths of
Lee-Wick vectors.

Consider the scattering of four
Lee-Wick gauge bosons. The scattering amplitude $\mathcal{M}$ is of
the form
\begin{equation}
\mathcal{M} = \epsilon(p_1)^\mu \epsilon(p_2)^\nu \epsilon(q_1)^\rho \epsilon(q_2)^\sigma \mathcal{M}_{\mu \nu \rho \sigma}
\end{equation}
where $\epsilon(p)$ is a polarization vector associated with momentum $p$ and $\mathcal{M}_{\mu \nu \rho \sigma}$ is a dimensionless quantity built from the Feynman rules of the theory. Since the Lagrangian in Lee-Wick form contains only dimension four operators, $\mathcal{M}_{\mu \nu \rho \sigma}$ does not grow at high energies with fixed scattering angle. Longitudinal polarization vectors, on the other hand, do grow at high energy: the longitudinal polarization vector associated with a particle of four-momentum $(E, p, 0, 0)$ is given by
\begin{equation}
\epsilon_L =  (p, E, 0, 0)/M.
\end{equation}
When the growth of longitudinal polarization vectors is taken into
account, we see that the amplitude $\mathcal{M}$ could grow as quickly
as $E^4$ at large energy. This kind of growth would be a disaster for
the theory.
%The fixed angle Froissart bound~\cite{Froissart:1961ux} for
%example, requires that amplitudes grow no faster than $s^{3/4}\log^{3/2}
%s$ for $t \neq 0$. 
In a theory of a massive vector boson with Lagrangian
\begin{equation}
\lag = -\frac{1}{2} \tr F_{\mu \nu} F^{\mu \nu} + M^2 \tr A_\mu A^\mu
\end{equation}
it is known that the amplitude describing four longitudinal vector boson
scattering grows as $E^2$. The possible $E^4$ growth is removed because to
a first approximation one can ignore the mass $M$ occurring in propagators
at high energies; then the Ward identity of gauge-invariant theories acts
to remove the largest growth. Nevertheless, the amplitudes still grow too
quickly with energy to be consistent with unitarity. Thus, the theory must
either become strongly coupled so that perturbative computations are
misleading, or new degrees of freedom must appear around the scale $M$
to restore unitarity. In non-Abelian Lee-Wick gauge theories, the Ward
identity is exact despite the presence of massive vector bosons. We
will show that the exact Ward identity prevents the amplitudes growing
on account of large polarization vectors.

\section{Ward Identities in the Higher Derivative Formulation}

We now turn to the proof that scattering amplitudes in Lee-Wick gauge
theory do not increase too quickly with energy. The key to the proof
is the following observation. For a massive vector boson with very large 
momentum $p$ with respect to some reference frame, the associated
longitudinal polarization vector is proportional to the momentum plus a
residual vector $\delta$ which does not grow with energy. For example,
if the momentum is $(E,p,0,0)$ then
\begin{equation}
\delta^\mu = (p-E, E-p, 0, 0)/M, \quad \lim_{p \rightarrow \infty} (p-E, E-p, 0, 0) = 0.
\end{equation}
The Ward identities (WI) provide us with non-perturbative information on
the structure of amplitudes with external momenta contracted into a leg.
We study the Ward identities in higher derivative gauge theories, and
show that they force amplitudes to vanish if an external momentum is
contracted in. This happens for ordinary gauge bosons as in normal gauge
theories, and also for the new Lee-Wick poles. Thus, because of gauge
invariance, we may always replace external longitudinal polarizations
with the residual vector $\delta^\mu$:
\begin{equation}
\epsilon_L(p) \sim \delta(p) = \epsilon_L - p /M .
\end{equation}
Since $\delta$ does not grow with energy, the high energy behaviour of
the amplitudes is given by dimensional analysis of the vertices in the
Lee-Wick form of the theory.

Let us now begin our study of the Ward identities. We will work in
background field gauge, so we derive the identities for one particle
irreducible (1PI) functions from the effective action\footnote{We assume
that the procedure used to define amplitudes that are unitary and contain
no cuts or poles from the Lee-Wick vector bosons does not violate gauge
invariance.}. Recall that this quantity is related to the 1PI functions by
\beq
\Gamma(A)=\sum_n\frac1{n!}\int\left(\prod_{i=1}^n d^4x_i\right)A^{a_1}_{\mu_1}(x_1)\cdots
A^{a_n}_{\mu_n}(x_n)\Gamma^{(n)a_1\cdots a_n}_{\phantom{(n)}\mu_1\cdots \mu_n}(x_1,\ldots,x_n) .
\eeq
Now, this is a gauge invariant quantity, so that 
\beq
\Gamma(A_\mu+\frac1{ig}D_\mu\omega)=\Gamma(A_\mu) 
\eeq
for infinitesimal $\omega^a(x)$. 
%That is
%\beq
%\int d^4x \frac{\delta\Gamma}{\delta A^a_\mu(x)}(D_\mu\omega)^a(x)=0
%\eeq
Consequently, the 1PI functions obey
\beq
\sum_n\frac1{(n-1)!}\int\left(\prod_{i=1}^n d^4x_i\right)(D_{\mu_1}\omega)^{a_1}(x_1)A^{a_2}_{\mu_2}(x_2)\cdots
A^{a_n}_{\mu_n}(x_n)\Gamma^{(n)a_1\cdots a_n}_{\phantom{(n)}\mu_1\cdots \mu_n}(x_1,\ldots,x_n)=0 .
\eeq
This implies the WIs, which are obtained by taking one functional
derivative with respect to $\omega$, $(n-1)$ functional derivatives with 
respect to $A$
and setting $\omega$ and $A$ to zero. Carrying out the algebra we obtain

\begin{equation}
\partial_{\mu_1} \Gamma^{(n)a_1\cdots a_n}_{\phantom{(n)}\mu_1\cdots \mu_n}(x_1,\ldots,x_n)=\\
g\sum_{i=2}^n\,\delta^{(4)}(x_1-x_i)\,f^{a_1a_ib}\,
\Gamma^{(n-1)a_1\widehat{a_2\cdots a_n}}_{\phantom{(n-1)}\mu_1\widehat{\mu_2\cdots \mu_n}}(x_1,\widehat{x_2,\ldots,x_n}) ,
\end{equation}
where the hat over the last $(n-2)$ entries of the lists indicates that 
the $i$-th entry should be removed.
Fourier transforming to momentum space, we find
\begin{equation}
p_{1\mu_1} \Gamma^{(n)a_1\cdots a_n}_{\phantom{(n)}\mu_1\cdots \mu_n}(p_1,\ldots,p_n)=\\
-ig\sum_{i=2}^n\,f^{a_1a_ib}\,
\Gamma^{(n-1)a_1\widehat{a_2\cdots a_n}}_{\phantom{(n-1)}\mu_1\widehat{\mu_2\cdots \mu_n}}(p_1+p_i,\widehat{p_2,\ldots,p_n}) ,
\label{eq:generalWard}
\end{equation}
where the momenta satisfy the condition $\sum p_i=0$. We now move on to
examine Ward identities explicitly in several cases.

\subsection{Uses of WIs I: The two point function}
Firstly, we apply Eq.~\eqref{eq:generalWard} to the case $n=2$. This is
simple since there is no $n=1$ 1PI function:
\beq
\label{WI2}
p^\mu \Gamma^{(2)ab}_{\phantom{(2)}\mu\nu}(p)=0 ,
\eeq
which implies
\beq
\label{two-point}
\Gamma^{(2)ab}_{\phantom{(2)}\mu\nu}(p)=-i\delta^{ab}(p^2g_{\mu\nu}-p_\mu p_\nu)\Pi(p^2) .
\eeq
More precisely, the WI does not determine the dependence on color indices,
but one can easily derive that separately. Note that in the higher
derivative theory, any zero of $\Pi(p^2)$ corresponds to an on-shell
degree of freedom, unlike in ordinary gauge theories for which $p^2=0$
is the on-shell condition.

\subsection{Uses of WIs II: The three point function}
The case $n=3$ is, explicitly, 
\beq
k^\mu
\Gamma^{(3)abc}_{\phantom{(3)}\mu\nu\lambda}(k,p,q)=
-igf^{abe}\Gamma^{(2)ec}_{\phantom{(2)}\nu\lambda}(k+p)-igf^{ace}\Gamma^{(2)eb}_{\phantom{(2)}\lambda\nu}(k+q) .
\label{eq:WI3}
\eeq
Now we can use the results in Eqs.~\eqref{WI2}-\eqref{two-point} on the right hand
side of Eq.~\eqref{eq:WI3}. It follows that if we put the momenta $p$ and $q$ on shell and
contract with either polarization vectors or momenta this
vanishes. Let us see how this works in some detail.

First, consider the case $q^2=p^2=0$. Then the right hand side of Eq.~\eqref{eq:WI3} is
\beq
-g f^{abc} (p_\lambda  p_\nu -q_\lambda  q_\nu )\Pi(0)
\eeq
If we now contract this with two polarization vectors,
$\epsilon^\nu(p)\epsilon^\lambda(q)$, or with the two remaining external
momenta $p^\nu q^\lambda$, the result is obviously zero. The same result
holds if we contract with one polarization and one momentum, and use
the condition $q^2=p^2=0$.

The case with $p^2=M^2$ and $q^2=0$ has the right hand side
\beq
g f^{abc}\left[(M^2g_{\lambda \nu}-p_\lambda  p_\nu)\Pi(M^2) +q_\lambda  q_\nu \Pi(0)\right]= g f^{abc} q_\lambda  q_\nu \Pi(0)
\eeq
and this obviously vanishes if one contracts with $\epsilon^\lambda(q)$ or $q^\lambda
$. Finally, the case with $p^2=q^2=M^2$ is trivial because $\Pi(M^2) = 0$.

%I have verified these results at tree level. They hold even including the
%$F^3$ term. There is nothing in this derivation that the $F^3$ term
%would invalidate.

\subsection{Uses of WIs III: The four point function and gauge invariance
  of the $S$-matrix}
In this case we find an identity which has a less obvious interpretation:
\begin{multline}
\label{WI4}
k^\mu 
\Gamma^{(4)abcd}_{\phantom{(4)}\mu\nu\lambda\sigma }(k,p,q,r)=-ig\left[
f^{abe}\Gamma^{(3)ecd}_{\phantom{(3)}\nu\lambda\sigma
}(k+p,q,r)\right.\\
\left.+f^{ace}\Gamma^{(3)bed}_{\phantom{(3)}\nu\lambda\sigma }(p,k+q,r)+
f^{ade}\Gamma^{(3)bce}_{\phantom{(3)}\nu\lambda\sigma }(p,q,k+r)\right] .
\end{multline}
The key to understanding the use of this identity is to write the
scattering amplitude, which is the sum of $\Gamma^{(4)}$ plus three more
terms corresponding to the $s$, $t$ and $u$ channel exchanges of a
gauge boson between $\Gamma^{(3)}$ vertices. What we will show is that
when we contract those with $k^\mu$, put the other particles on shell,
and also contract with external polarizations or momenta, then the sum
precisely cancels the three terms of $k^\mu \Gamma^{(4)}_{\mu \nu \lambda
\sigma}$ above.

For example, the $s$-channel exchange amplitude is 
\beq
-\Gamma^{(3)abe}_{\phantom{(3)}\mu \nu\rho}(k,p,-k-p) G^{(2)eh}_{\phantom{(2)}\rho\eta}(k+p)
\Gamma^{(3)cdh}_{\phantom{(3)}\lambda\sigma\eta }(q,r,k+p) 
\eeq
where we have used the same combination of momenta and indices as in
the four point functions above. Also we have introduced the full propagator
$G^{(2)}$. This, of course only makes sense if it is gauge fixed but the
WIs of the  previous section show that the gauge fixing term gives no
contribution once the the external legs are put on-shell and contracted
with polarizations or momenta. Now, contract this with $k^\mu$ and use
\beq
k^\mu \Gamma^{(3)abe}_{\phantom{(3)}\mu \nu\rho}(k,p,-k-p) 
= ig\left[ f^{hba}\Gamma^{(2)he}_{\phantom{(2)}\nu\rho}(k+p)+ 
f^{hea}\Gamma^{(2)hb}_{\phantom{(2)}\rho \nu}(-p)\right]
\eeq
We can ignore the second term, since $\Gamma^{(2)hb}_{\phantom{(2)}\rho
  \nu}(-p)\epsilon^\nu (p)=0$ and $\Gamma^{(2)hb}_{\phantom{(2)}\rho
  \nu}(-p)p^\nu=0$. The first term remains. Let us contract it
with the propagator:
\beq
ig f^{hba}\Gamma^{(2)he}_{\phantom{(2)}\nu\rho}(k+p)
G^{(2)el}_{\phantom{(2)}\rho\eta}(k+p)
= ig f^{lba} \left(g_{\nu\eta}-\frac{(k+p)_\nu (k+p)_\eta}{ (k+p)^2}\right) .
\eeq
We have used the fact that $\Gamma^{(2)}$ is the inverse of $G^{(2)}$,
but only on the space projected out by the former. Hence the $s$-channel
exchange contracted with $k^\mu$ gives
\beq
-ig f^{hba} \left(g_{\nu\eta}-\frac{(k+p)_\nu (k+p)_\eta}{ (k+p)^2}\right)
  \Gamma^{(3)cdh}_{\phantom{(3)}\lambda\sigma\eta }(q,r,k+p)=
-ig f^{hba}  \Gamma^{(3)cdh}_{\phantom{(3)}\lambda\sigma\nu }(q,r,k+p)
\eeq
up to terms that vanish when external legs go on shell and are contracted with
polarization vectors or momenta. This term cancels the first term in
\eqref{WI4}. The second and third terms in \eqref{WI4}  are canceled
by the $t$ and $u$-channel exchanges. Thus, we see that the Ward identity
forces four particle scattering amplitudes to vanish for on-shell external
particles when one of more external momenta are contracted into the legs.
This removes the growth of scattering amplitudes associated with large
polarization vectors.

\section{Successes and Failures of Dimensional Analysis}

We have now seen that the growth of longitudinal polarization vectors
in the higher derivative theory is not important in scattering amplitudes.
However, the amplitudes could still grow if the uncontracted amplitude
$\mathcal{M}_{\mu \nu \rho \sigma}$ grows. In fact, in the higher derivative
theory of Eq.~\eqref{eq:gbHDlag} one might expect this amplitude to grow.
The reason is that the four particle scattering amplitude must be
dimensionless. However, there are vertices in this theory which are
proportional to $1 / M^2$. Thus, by dimensional analysis (DA)
the rest of the interaction must have mass dimension 2, so that terms
like $E^2$ are allowed. These terms would eventually lead to violation
of perturbative unitarity.

However, the Lee-Wick description of the theory shown in
Eq.~\eqref{eq:gbLWlag} is equivalent to the higher derivative
formulation. In the Lee-Wick description, only operators of dimension
four are present. Thus, the four vector boson vertex, for example,
is dimensionless. Now DA indicates that at high energies, $E \gg M$,
the rest of the interaction consists of dimensionless ratios formed from
the momenta in the problem, so that the uncontracted amplitude does not
grow at high energies with fixed non-zero scattering angle. 
% Of course,
%the scattering amplitudes in the higher derivative and Lee-Wick forms
%of the theory are related since, in path integral terms,
%\begin{multline}
%\int D \hat A \, \hat A_\mu \hat A_\nu \hat A_\rho \hat A_\sigma \exp
%\left( \int d^4 x \lag_{\mathrm{hd}} \right) \\
%= \int D A \, D \tilde A \, (A_\mu + \tilde A_\mu) (A_\nu + \tilde A_\nu)
%(A_\rho + \tilde A_\rho) (A_\sigma + \tilde A_\sigma) \exp
%\left( \int d^4 x \lag_{\mathrm{LW}} \right) .
%\end{multline}
Of course, the on-shell scattering amplitudes in the higher derivative
and Lee-Wick forms of the theory are the same.
Since we can compute the scattering amplitude appropriate for
the higher derivative theory from the Lee-Wick form, we conclude
that the full amplitude does not grow with energy. Now when we
contract in the polarization vectors, we see that their growth is also
unimportant. Putting it all together, we find that the on-shell scattering
amplitudes cannot grow at high energies.

The Lagrangian given in Eq.~\eqref{eq:gbHDlag} is not the most general
Lagrangian including dimension six operators. One could also add a term
\begin{equation}
\Delta \lag = \frac{i \gamma g}{M^2} \tr \left( \hat F_{\lambda \mu} [ \hat F^{\lambda \nu} , \hat F^\mu{}_\nu] \right).
\end{equation}
It is still possible to construct a Lee-Wick Lagrangian to describe
the theory including this operator in the same way as described
in Section~\ref{sec:review}.
However, the resulting Lagrangian contains dimension six
operators and so we expect the amplitudes to grow with energy. We have
confirmed that this is the case for the operator in $\Delta \lag$ by
explicit calculation.  Thus, internal consistency of Lee-Wick gauge
theories requires the special
choice of higher dimension operator shown in Eq.~\eqref{eq:gbHDlag}.

\section{Explicit Calculations of Scattering Amplitudes}

We have demonstrated that the amplitudes for vector
boson scattering in Lee-Wick theory do not grow with energy. 
%Since the theory can be
%formulated with auxilliary fields and operators of dimension less than
%or equal to four the two-two scattering amplitudes can only grow due
%to the longitudinal polarization vectors. Using the higher derivative
%formulation of the theory where the massive gauge bosons are associated
%with poles in the gauge field propagator we used gauge invariance to
%argue that their on-shell scattering amplitudes vanish when any of the
%LW-fields polarization vectors are replaced by their momentum. The
%combination of these results ensures that the scattering amplitudes
%for massive LW-vector bosons does not grow with energy resulting in a
%violation of perturbative unitarity.
%
%In the LW standard model there are massive non-Abelian gauge bosons associated with the color $SU(3)$ group and the weak $SU(2)$. 
Since our argument for acceptable high energy behavior uses a combination of the two formulations of the theory it is worth presenting the results of some explicit calculations that support our conclusions. For definiteness we 
consider in this section the scattering  $\tilde W^1(p_1) \tilde W^2(p_2) \rightarrow \tilde W^2(q_1) \tilde W^1(q_2)$ of $SU(2)$ LW-gauge bosons. Here the superscripts denote the adjoint gauge indices associated with the particles being scattered. We work in the center of mass frame and take the incoming particles to have energy $E$ and let $\theta$ be the angle between the momenta  ${\bf p}_1$ and ${\bf q}_1$. 

For the case where all the LW- bosons are longitudinally polarized we find that the leading behaviour of the scattering amplitude at high energy $E$ and fixed scattering angle $\theta$ is
\begin{equation}
\mathcal{M}(LL \rightarrow LL) \simeq g^2 {1+{\rm cos}\theta \over (1-{\rm cos}\theta)^2}\left( M \over E \right)^2 ( (19-9{\rm cos}\theta){\rm cos}\theta+14),
\end{equation}
where $g$ is the $SU(2)$ gauge coupling, and the symbol $\simeq$ indicates that we have shown only the leading behaviour in an expansion in powers of $M/E$. We have performed the tree level calculation of this amplitude both in the higher derivative formulation of the theory and the formulation with LW-fields. In the LW-field formulation the particular values of the coefficients of the operators in Eq.~\eqref{eq:gbLWlag} encode the fact that the massive LW-vector bosons are associated with poles in a gauge field and are crucial for obtaining a scattering amplitude that does not grow with energy. If the Lagrange density,
\begin{equation}
\lag = - \frac{1}{2} \tr {\tilde F}_{\mu \nu}{\tilde  F}^{\mu \nu}+M^2 \tr{\tilde A}_{\mu}{\tilde A}^{\mu},
\end{equation}
where $\tilde F_{\mu \nu} = \partial_\mu \tilde A_\nu - \partial_\nu \tilde
A_\mu - i g [ \tilde A_\mu, \tilde A_\nu]$, was used to calculate this scattering amplitude it would grow proportional to $E^2$ resulting at high energies in a theory that is either strongly coupled or violates unitarity.

Next we consider the case where the final LW gauge boson with four-momentum $q_2$ is transversely polarized in the plane of the scattering. In that case the high energy behaviour of the scattering amplitude is
\begin{equation}
\mathcal{M}(LL \rightarrow TL) \simeq g^2 \cos \theta \sqrt{{1+{\rm cos}\theta \over 1-{\rm cos}\theta}} {M \over E}.
\end{equation}
Not all the amplitudes fall with increasing energy. For example, the scattering of two longitudinal LW-vector bosons into two that are transversely polarized in the plane of scattering does not grow at high
energies,
\begin{equation}
\mathcal{M}(LL \rightarrow TT)\simeq g^2 (1+{\rm cos}\theta) .
\end{equation}

\section{Concluding Remarks}

Non-Abelian Lee-Wick gauge theories contain massive vector bosons whose
scattering does not violate perturbative unitarity, unlike ordinary
gauge theories with mass terms. 
%It may be possible to build a theory
%without ghosts or instabilities and which still consistently describes
%a weakly coupled massive vector boson. It would be very interesting to
%find such a theory.
There still remains the question of whether Lee-Wick theories are
Lorentz invariant and unitary to all orders of perturbation theory, and of whether
this can be done while preserving gauge invariance. So
far, only specific examples have been studied~\cite{CLOP, Kuti,
Antoniadis:1986tu}. It would be very interesting to answer this question.

\acknowledgements

We thank Nima Arkani-Hamed for discussions that lead to this paper. The
work of BG, DOC and MBW was supported in part by the US Department
of Energy under contracts DE-FG03-97ER40546, DE-FG02-90ER40542 and
DE-FG03-92ER40701, respectively.

\end{document}